\documentstyle[12pt,epsf,twoside,fleqn,espcrc1]{article}

\newcommand{\be}{\begin{eqnarray}}
\newcommand{\ee}{\end{eqnarray}}

\newcommand{\AmS}{{\protect\the\textfont2
  A\kern-.1667em\lower.5ex\hbox{M}\kern-.125emS}}

\title{Heavy Ion Physics at the LHC\thanks{This work was supported
in part by the Director, Office of Energy Research, Division of Nuclear Physics
of the Office of High Energy and Nuclear Physics of the U. S.
Department of Energy under Contract No. DE-AC03-76SF0098.}}

\author{R. Vogt\address{Lawrence 
Berkeley National Laboratory, Berkeley, CA 
USA}$^,$\address{Physics Department, 
University of California, Davis, CA 
USA}}

\begin{document}

\maketitle

\begin{abstract}
The ion-ion center of mass energies at the LHC will 
exceed that at RHIC by nearly a factor of 30, providing exciting 
opportunities for addressing unique physics issues in a completely new 
energy domain. Some highlights of this new physics domain 
are presented here.  We briefly describe how
these collisions will provide new insights into the high density, low momentum
gluon content of the nucleus expected to dominate the dynamics of the early
state of the system.  We then discuss
how the dense initial state of the nucleus affects the lifetime and
temperature of the produced system.  Finally, we explain how the high energy
domain of the LHC allows abundant production of `rare' processes,
hard probes calculable in perturbative quantum chromodynamics, QCD.  At the
LHC, high momentum jets and $b \overline b$ bound states, the $\Upsilon$
family, will be produced with high statistics for the first time in heavy ion
collisions. 
\end{abstract}\\

\section{QUANTIFYING THE INITIAL STATE}

An accelerated nucleus may be envisioned as a lattice of valence quarks
surrounded by sea quark and gluon fields.  Although these sea quarks and gluons
carry only a small fraction, $x$, of the total nucleon momentum, their density
is very high, especially for the gluons.  As the energy of the ion beam is
increased, the lowest $x$ values probed decreases while the density increases.
Typical $x$ values of partons produced at midrapidity, $y=0$, with transverse
momentum of $p_T \approx 2$ GeV/$c$ at RHIC are $x \sim 2p_T/\sqrt{s_{NN}} \sim
0.02$ where the gluon density is not yet very high.  However, the factor of 30
increase in energy between RHIC and the LHC decreases the $x$ values
correspondingly to $\sim 6.7 \times 10^{-4}$ where the gluon density is quite
high.  Expanding the rapidity coverage to the forward region further reduces
the $x$ values probed while increasing the gluon density still more.  

At these high gluon
densities where $x$ is low and the 4-momentum transfer squared, $Q^2$,
is moderate, the $Q^2$ evolution of the gluon densities can no longer be
described by standard, linear evolution in $Q^2$.  Instead, in the regime
$1.5 \leq Q^2 \leq 10$ GeV$^2$ and $10^{-5} \leq x \leq 5 \times 10^{-3}$,
nonlinear evolution of the parton densities dominates.  The gluon wavelengths
are long enough that they overlap each other and begin to interact, leading to
terms with squared gluon densities at low $x$.  At still smaller $x$,
$x \leq 10^{-5}$, the gluon density is so large that it saturates the available
phase space and the dynamics of the interaction may be described in terms of
classical color fields.  

Although the numbers given here are relevant for $pp$
collisions where the gluons in a single nucleon begin to overlap each other, 
since the nonlinear growth of the gluon density depends on the
transverse size of the system, these effects, as well as the subsequent
saturation physics, may be expected to set in at higher $x$ for nuclei than for
free nucleons.  Thus the initial state of the system and the gross properties
of the system may be calculable in perturbative QCD.

The LHC energy domain is clearly in the regime where small $x$ effects and
departures from linear $Q^2$ evolution of the parton distributions will be
prominent.  Systematic studies of proton-proton, $pp$, and proton (or
deuteron)-nucleus, $p$(d)$A$, collisions can yield much exciting information
about the initial state of nucleus-nucleus, $AA$, collisions.  The $pp$
collisions will be at the highest
energy, $\sqrt{s} = 14$ TeV, and therefore the lowest $x$.  These collisions
can study the nonlinear evolution regime and probe the onset of saturation in
the proton.  Studies of $p$Pb or dPb collisions at 8.8 and 6.2 TeV/nucleon
respectively, can elucidate the difference between the saturation regimes of
the proton and the nucleus, providing the baseline nuclear parton distributions
necessary to fully understand Pb+Pb collisions at $\sqrt{s_{NN}} = 5.5$ 
TeV/nucleon.

Another important characteristic of the initial state is its baryon number.
One expects that,
during the course of the collision, the valence quarks will be swept away from
the center of the reaction zone, taking the total baryon number with them to
the fragmentation regions.  As a result, protons and antiprotons should be
produced in equal abundance in the central region
at sufficiently high energies.  A manifestation
of this is the measurement of the antiproton-to-proton ratio, $\overline p/p$,
which has been shown to grow with energy.  The most recent measurements at
RHIC indicate $\overline p/p \sim 0.65$ in $\sqrt{s_{NN}} = 130$ GeV Au+Au
collisions \cite{pbaroverp1}, growing to $\sim 0.77$ for $\sqrt{s_{NN}} = 200$
GeV \cite{brahmspbop,pbaroverp2}.  The increase of this ratio with energy
strongly suggests that, while a baryon free regime has not yet been reached
at RHIC, the chances for reaching this regime at the LHC are quite good.

Thus the initial state of nuclear collisions at the LHC can be characterized by
near zero baryon density and high initial gluon densities that may be 
described in terms of classical color fields.  The initial state may be most
cleanly studied in $pA$ interactions where the multiplicities are still small.
In addition, the best probes of such matter are those which do not interact
strongly such as real and virtual photons and massive gauge bosons, the $W^\pm$
and $Z^0$.  

\begin{figure}[htb]
\setlength{\epsfxsize=0.95\textwidth}
\setlength{\epsfysize=0.5\textheight}
\leftline{\epsffile{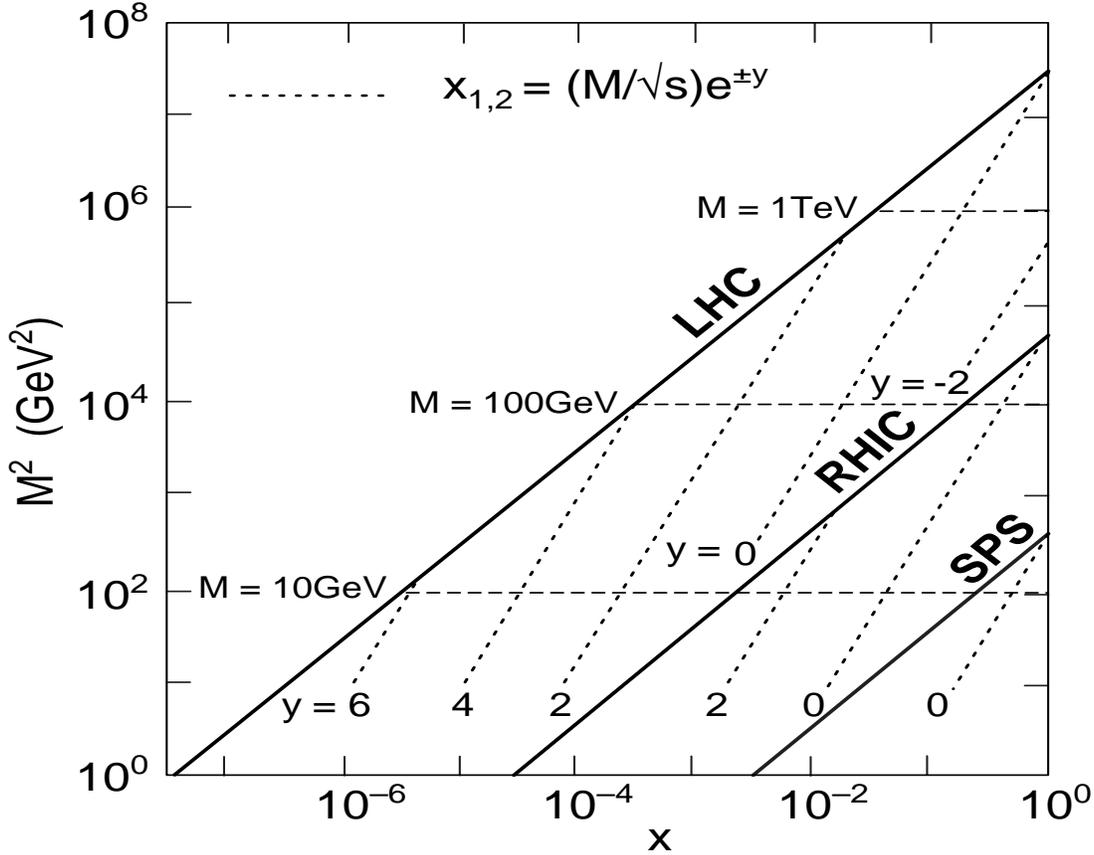}}
\caption[]{The $Q^2 = M^2 \geq 1$ GeV$^2$ reach as a function of $x$ for the 
SPS, RHIC and the LHC.  Lines of constant rapidity are indicated for each
machine.}
\end{figure}

We know from low energy deep-inelastic scattering that the nuclear
parton distributions are modified relative to those of the free proton.
However, the measurements are not very sensitive to the gluon distributions and
the low $x$ measurements are limited to rather small $Q^2$, $Q^2 \ll 1$
GeV$^2$.  At the LHC however, the low $x$ region can be effectively probed in
the perturbative regime, up to $Q^2 \geq m_Z^2$.  Massive gauge boson
production proceeds predominantly through the $q \overline q$ channel rather
than by gluons.  Thus gauge boson production
is an efficient probe of the high $Q^2$ quark and antiquark distributions. 
At RHIC, only the upper $pp$ energy is  large enough for statistically
significant studies of $W^\pm$ and $Z^0$  production since the rates in the
lower energy $AA$ collisions are too small for meaningful 
measurements. High-statistics studies of the nuclear quark and antiquark 
densities  at large momentum transfer, $Q^2$, through measurements of $W^\pm$ 
and $Z^0$ production would thus be unique to the LHC~\cite{VogZ}.  

Direct
photon production, on the other hand, is dominated by $qg \rightarrow q\gamma$
and can efficiently study the nuclear gluon distributions.  Open charm
production is another gluon-dominated process that helps determine the nuclear
gluon distribution.  We note that the effects of modified parton density
evolution such as nonlinear and saturation effects should manifest themselves
most strongly at low $Q^2$.  Thus charm, with its relatively low mass, is
also an important probe of the small $x$, low $Q^2$ regime \cite{ekv2}.
Since modifications of the parton densities in nuclei have been observed when
neither $x$ nor $Q^2$ is small, measurements of the nuclear parton
distributions at the LHC can help disentangle the different regimes of parton
evolution.

\section{THE FINAL STATE:  HOTTER AND LONGER LIVED}

Heavy ion collisions at LHC energies will explore regions of energy and
particle density significantly beyond those reachable at RHIC.
The energy density of the thermalized matter created at the LHC is 
estimated to be 20 times higher than at RHIC,
implying an initial temperature, $T_0$, nearly a factor of two higher than at
RHIC \cite{ivanqm}.  The higher densities of the produced partons results in
more rapid thermalization.  Consequently, the time spent in the quark-gluon
plasma phase, the difference between the plasma lifetime and the thermalization
time, increases by almost a factor of three \cite{ivanqm}.  Thus the hot, 
dense systems created in heavy 
ion collisions at the LHC spend more time in a purely partonic 
state. The longer lifetime of the quark-gluon plasma state
widens the time window available to probe it experimentally, as seen in
Table~\ref{ivan}. 

\begin{table}[htb]
\caption[]{The initial QGP production time, lifetime, initial temperature and
energy density for the maximum energy and mass systems at the SPS, RHIC and the
LHC.  From Ref.~\protect\cite{ivanqm}.}
\label{ivan}
\begin{center}
\begin{tabular}{|cccccc|} \hline
System & $\sqrt{s_{NN}}$ (GeV) & $\tau_0$ (fm) & $\tau_{\rm tot}$ (fm) & 
$T_0$ (MeV) & $\epsilon_0$ (GeV/fm$^3$)  \\ \hline 
SPS (Pb+Pb)  & 17   & 0.8 & $1.4 - 2$ & $210 - 240$ & $1.5 - 2.5$ \\
RHIC (Au+Au) & 200  & 0.6 & $6 - 7$   & $380 - 400$ & $14 - 20$   \\
LHC (Pb+Pb)  & 5500 & 0.2 & $18 - 23$ & $710 - 850$ & $190 - 400$ \\
\hline 
\end{tabular}
\end{center}
\end{table}

\section{ABUNDANT HARD PROBES}

Semi-hard and hard parton processes dominate particle production at the LHC.
These hard probes~\cite{XW1}, such as high-$p_T$ jets and photons, 
quarkonia, and $W^\pm$ or $Z^0$ bosons, are characterized by the $Q^2$ 
necessary for their production. At the high collision energies of the 
LHC, the cross sections for processes with $Q^2 > $\,(50\,GeV)$^2$ 
are large enough for detailed systematic
studies.

To better illustrate the large rates available for hard probes at the LHC,
in Table~\ref{introtab} we present the minimum bias
jet and gauge boson rates in the region $|\eta| \leq 2.4$ as well as the total
$Q \overline Q$ and quarkonium rates calculated in perturbative QCD in Pb+Pb
collisions at $\sqrt{s_{NN}} = 5.5$ TeV and $p$Pb collisions at $\sqrt{s_{NN}}
= 8.8$ TeV.  The results are given for a $10^6$~s LHC run in each case.
We have assumed a luminosity of $5 \times
10^{26}$~cm$^{-2}$s$^{-1}$ for Pb+Pb \cite{brandt}
and a maximum $p$Pb luminosity of $1.4
\times 10^{30}$~cm$^{-2}$s$^{-1}$ \cite{yellow_report_pa}.  Conventional
nuclear shadowing, the only nuclear effect included, is typically rather small
for jets and gauge bosons but can be large for heavy quarks and quarkonium.
The quarkonium rates include the branching ratios to lepton pairs.

\begin{table}[htb]
\caption[]{The yield of hard probes in a $10^6$~s LHC run.}
\begin{center}
\begin{tabular}{|ccc||cc|} \hline
& \multicolumn{2}{c||}{Pb+Pb} & \multicolumn{2}{c|}{$p$Pb} \\
& \multicolumn{2}{c||}{$\sqrt{s_{NN}} = 5.5$ TeV} & 
\multicolumn{2}{c|}{$\sqrt{s_{NN}} = 8.8$ TeV} \\
& \multicolumn{2}{c||}{${\cal L} = 5\times
10^{26}$~cm$^{-2}$s$^{-1}$} & \multicolumn{2}{c|}{${\cal L} = 
1.4\times 10^{30}$~cm$^{-2}$s$^{-1}$} \\ \hline
Process & Yield/$10^6$~s & Ref. & Yield/$10^6$~s & Ref. \\ \hline
\multicolumn{5}{c}{$|\eta|\leq 2.4$} \\ \hline
jet($p_T > 50$ GeV)                 & $2.2 \times 10^7$    &
\cite{yellow_report_jets} & $1.5 \times 10^{10}$ & \cite{yellow_report_pa}  \\
jet($p_T > 250$ GeV)                & $2.2 \times 10^3$    &
\cite{yellow_report_jets} & $5.2 \times 10^6$    & \cite{yellow_report_pa}  \\
$Z^0$                               & $3.2 \times 10^5$    &
\cite{VogZ}               & $6.8 \times 10^6$    & \cite{yellow_report_pa}  \\
$W^+$                               & $5.0 \times 10^5$    &
\cite{VogZ}               & $1.1 \times 10^7$    & \cite{yellow_report_pa}  \\
$W^-$                               & $5.3 \times 10^5$    &
\cite{VogZ}               & $1.1 \times 10^7$    & \cite{yellow_report_pa}  \\
\hline
\multicolumn{5}{c}{all phase space} \\ \hline
$c \overline c$                     & $9.0 \times 10^{10}$ &
\cite{RVpriv}             & $2.0 \times 10^{12}$ & \cite{RVpriv}            \\
$b \overline b$                     & $3.6 \times 10^9$    &
\cite{RVpriv}             & $8.2 \times 10^{10}$ & \cite{RVpriv}            \\
$J/\psi \rightarrow \mu^+\mu^-$     & $2.4 \times 10^7$    &
\cite{yellow_report_qqb}  & $5.5 \times 10^8$    & \cite{yellow_report_qqb} \\
$\Upsilon \rightarrow \mu^+\mu^-$   & $1.5 \times 10^5$    &
\cite{yellow_report_qqb}  & $3.5 \times 10^6$    & \cite{yellow_report_qqb} \\
$\Upsilon' \rightarrow \mu^+\mu^-$  & $3.7 \times 10^4$    &
\cite{yellow_report_qqb}  & $8.4 \times 10^5$    & \cite{yellow_report_qqb} \\
$\Upsilon'' \rightarrow \mu^+\mu^-$ & $2.2 \times 10^4$    &
\cite{yellow_report_qqb}  & $5.2 \times 10^5$    & \cite{yellow_report_qqb} \\
\hline 
\end{tabular}
\label{introtab}
\end{center}
\end{table}

\section{MODIFICATION OF JETS IN DENSE MATTER}

High $p_T$ quark and gluon jets can be used to study the
the hot medium produced after the collision.
The large $Q^2$ of these jets causes them to materialize immediately
after the collision.  They are then
embedded in and propagate through the dense environment 
as it forms and evolves. Through their
interactions with the medium, they measure its properties
and are thus sensitive to the formation of a 
quark-gluon plasma. Large transverse momentum probes are easily 
isolated experimentally from the soft particle background of the collision. 
Their high $p_T$ ensures
that the medium effects are perturbatively calculable, 
strengthening their usefulness as quantitative diagnostic tools. 
At the LHC, the production rates for jets pairs with $p_T>50$ GeV for a single
jet are several orders of magnitude larger than at RHIC.  Indeed, more than 10
jet pairs are produced every second in Pb+Pb collisions at the LHC, as shown
in the previous section.  Thus high statistics systematic
studies are possible in a clean kinematic regime, far 
beyond the limits of RHIC.

Jet pairs should be produced back to back.  However, one or both of the jets
can be modified by the medium.  Interactions with the medium can reduce the
jet energy and change its direction.  In fact, some jets may lose so much
of their initial energy that their energy is dissipated and they no longer
appear as individual jets.  The appearance of monojets was suggested as an
early measure of `jet quenching' due to energy loss \cite{plumer}.  At RHIC,
evidence for such an effect has been measured through leading hadron
correlations.  In peripheral heavy ion collisions, two high $p_T$ hadrons
are detected back to back, similar to the measurements of $pp$ interactions
at the same energy.  However, in central collisions, the opposite side hadron
has disappeared and the azimuthal correlation is similar to that of the
underlying soft hadrons in the event \cite{star_jet}.  This jet suppression
is absent in dAu collisions at RHIC, strongly suggesting that the suppression
is a final-state effect arising from the dense medium.  Such leading hadron
measurements are also possible at the LHC.  In addition, at the higher energy,
the measurements can be performed for full jets rather than leading
hadrons and extended to significantly higher $p_T$. 

The dijet probe is nontrivial because the original energy of both jets is
unknown. It would be preferable to have a method of tagging jets of known
energy to measure the energy loss \cite{Wang}.
Quark jets of known energy can be produced in reactions such as 
$g q \rightarrow q \gamma$ or $g q \rightarrow q  Z^0$.  In these cases,
the energy is known since, to tree level, the quark jet is produced with
transverse momentum equal and opposite to that of the gauge boson which is
unaffected by the presence of the medium.  While the reaction $gq \rightarrow q
Z^0$ is a small contribution to the total $Z^0$ yield \cite{VogZ}, it is a more
distinctive signature since the $Z^0$ is free from the high background of
hadronic decays contributing to the direct photon spectrum \cite{cmsNote}.
At the LHC,  the $Z^0+$jet yield is large enough to extract physics signals.
Any energy loss suffered by the jet can thus be more
cleanly identified than in the dijet channel.

The measured energy loss yields the opacity of the medium:  the
product of the interaction cross section between the hard probe and the
partonic medium with its density~\cite{Mik,Wiedemann}.
In kinematic regions where the rescattering cross section can be 
reliably calculated, the opacity provides access to the parton density after 
the collision, determining how the medium is affected by gluon 
saturation.

In addition to the dijet and jet$+\gamma,Z^0$ measurements, other measurements
of high $p_T$ jets and hadrons may also be important for quantifying the energy
loss.  We discuss three of these below: jet fragmentation, jet shapes and high
$p_T$ heavy quarks.

While the initial hard scattering can be described in terms of 
proton and nuclear parton distribution functions, the process by which the
produced partons become the measured final state hadrons are characterized by
fragmentation functions.  These fragmentation functions are typically studied
in $e^+e^-$ collisions and then applied to other processes since they are
assumed to be universal.  However, because the parton distribution functions
are known to be modified by the presence of the nuclear medium, one might also
expect the fragmentation functions to be modified in the medium as well.
The medium in question could either be cold matter, as studied in $eA$
collisions at HERA \cite{hermes}, 
or hot and dense matter, produced in heavy ion
collisions.  Medium-modified fragmentation functions have been calculated by
averaging the vacuum fragmentation function over the parton energy loss in the
medium, weighted by the probability for the parton to lose energy.  The jet
suppression effects seen at RHIC are compatible with jet quenching
\cite{yellow_report_jets}.  Extrapolating from RHIC to the LHC based on the
expected gluon multiplicity at each energy, one can expect similar suppression
at $p_T > 50$ GeV in Pb+Pb collisions at the LHC.  

One might also tend to expect that energy loss would modify the final shape
of the produced jet. Recent calculations have found that such modifications are
small. While energy loss does not strongly modify the jet shape, it can
significantly change the multiplicity distribution inside the jet cone,
increasing the multiplicity by more than a factor of two
\cite{yellow_report_jets}.   A measurement of the $p_T$ distribution of
produced hadrons relative to the jet axis may be very sensitive to the $p_T$
broadening of the parton shower.  Such measurements are only possible at the
LHC where fully formed high $p_T$ jets are observable.

Heavy quark production is copious at the LHC.  These quarks tend to be produced
in the early stages of the collision because their mass is typically much
larger than the temperature of the medium.  While in the medium, these quarks 
can also undergo energy loss.  Recent calculations indicate that this loss
may be rather small.  However, the loss might be observable in either the
dilepton or single lepton channels at large invariant mass or large $p_T$
respectively, particularly for the $b$ quarks \cite{yellow_report_jets}.

\section{$\Upsilon$ SUPPRESSION AS A QUARK-GLUON PLASMA PROBE}

One of the proposed signatures of the QCD phase transition is the suppression
of quarkonium production \cite{MS,KMS}.  Suppression of the
$J/\psi$ and $\psi^\prime$ has been observed in
nucleus-nucleus collisions at the CERN SPS \cite{na50}.  
In a plasma, the suppression
occurs due to the shielding of the $c \overline c$ binding potential by color
screening, leading to the breakup of the resonance.  The $c \overline c$ and $b
\overline b$ resonances have smaller radii than light-quark hadrons and
therefore need higher temperatures to break up.
Because the $\Upsilon$ is much smaller
than the $c \overline c$ and other $b \overline b$ resonances, a much higher
temperature
is needed to dissociate the $\Upsilon$ \cite{KMS}.  Therefore it
was previously assumed that the $\Upsilon$ would not be suppressed by QGP
production \cite{KMS,KS}.  

In view of the high initial temperature 
of a gluon-dominated minijet plasma, $T\sim 0.9-1$ GeV \cite{EKinit}, it was
shown that, depending upon the properties of the plasma, the
$\Upsilon$ could be suppressed, providing a valuable tool to determine the
initial state of the system and the characteristics of the plasma \cite{gunv}.
With such high temperatures, strong plasma suppression might be expected.  
Unfortunately the short equilibration time 
of the minijet system correspondingly
reduces the plasma lifetime in the scaling expansion, causing the minijet
plasma to be too short-lived to produce quarkonium suppression in some cases.
Alternatively, the initial conditions could be dominated by kinetic
equilibration processes \cite{XKSW} with a correspondingly longer
equilibration time, $t_0 \sim 0.5-0.7$ fm.    
Because the equilibration time of the parton gas is longer than
that obtained from the minijet initial conditions, the time the system spends
above the breakup temperature is also longer, leading to stronger
suppression even though $T_0$ is lower.  

\begin{table}[htbp]
\caption[]{LHC values of $t_D$, and $p_{T {\rm m}}$, 
Ref.\ \protect\cite{cmsNote}.} 
\label{tdpttablelhc}
{\footnotesize
\begin{center}
%\begin{tabular}{|c|c|c|c|c|} \hline
\begin{tabular}{ccccc} %\hline
%\multicolumn{5}{c}{LHC} \\ \hline\hline
& \multicolumn{2}{c}{$\mu \propto gT$, $n_f = 3$, $T_c = 170$ MeV}  & 
\multicolumn{2}{c}{$\mu = 4T$, $T_c = 260$ MeV} \\ \hline
%\multicolumn{5}{c}{parton gas} \\ \hline
& \multicolumn{4}{c}{parton gas, $T_0=820$ MeV, $t_0=0.5$ fm} \\ \hline
 & $t_D$ (fm) & $p_{T {\rm m}}$ (GeV) &  $t_D$ 
(fm) & $p_{T {\rm m}}$ (GeV) \\ \hline
$\Upsilon$   & -     & 0     & 4.6   & 56.53    \\ %\hline
$\Upsilon'$  & 4.79  & 23.16 & 15.69 & 81.98 \\ %\hline
$\chi_b$     & 8.90  & 32.42 & 15.69 & 58.9\\ \hline
\multicolumn{5}{c}{minijet plasma, no shadowing} \\ \hline
& \multicolumn{2}{c}{$T_0=820$ MeV, $t_0=0.1$ fm}  & 
\multicolumn{2}{c}{$T_0=1.05$ GeV, $t_0=0.1$ fm} \\ \hline
 & $t_D$ (fm) & $p_{T {\rm m}}$ (GeV) &  $t_D$ 
(fm) & $p_{T {\rm m}}$ (GeV) \\ \hline
$\Upsilon$ & - & 0  & 1.94 & 22.2 \\ %\hline
$\Upsilon'$ & - & 0 & 6.59 & 33.2 \\ %\hline
$\chi_b$  & - & 0  & 6.59 & 23.05\\ \hline
\multicolumn{5}{c}{minijet plasma, HPC shadowing\protect\cite{HPCshad}} 
\\ \hline
& \multicolumn{2}{c}{$T_0=699$ MeV, $t_0=0.1$ fm}  & 
\multicolumn{2}{c}{$T_0=897$ MeV, $t_0=0.1$ fm} \\ \hline
 & $t_D$ (fm) & $p_{T {\rm m}}$ (GeV) &  $t_D$ 
(fm) & $p_{T {\rm m}}$ (GeV) \\ \hline
$\Upsilon$ & - & 0  & 1.21 & 11.7\\ %\hline
$\Upsilon'$ & - & 0 & 4.11 & 19.2 \\ %\hline
$\chi_b$  & - & 0  & 4.11 & 12.1\\ \hline
\end{tabular}
\end{center}
}
\end{table}

The time at which the temperature drops below $T_D$ and the state can no longer
be suppressed, $t_D = t_0(T_0/T_D)^3$, and the maximum quarkonium $p_T$ for
which the resonance is suppressed, $p_{T,{\rm m}} = M \sqrt{(t_D/\tau_F)^2
-1}$, are given in Table~\ref{tdpttablelhc} 
for $\mu(T) \propto gT$ with $n_f = 3$
and $\mu(T) = 4T$, SU(3) plasma with $T_c = 260$ MeV using 
the parton gas and minijet
initial conditions.  Results for the minijet
initial conditions are given for the GRV 94 LO parton densities both without
shadowing and HPC shadowing \cite{HPCshad}, resulting in 
the lowest temperatures obtained with shadowing \cite{cmsNote}.
Note that the reduction of the initial temperature due to shadowing
significantly reduces the $p_T$ range of the suppression.  However, this result
can be distinguished from a case with no significant shadowing and a plasma
with a smaller spatial extent \cite{gunv}. 

A high statistics study of quarkonium production ratios such as $\psi'/\psi$
and $\Upsilon'/\Upsilon$ 
as a function of $p_T$ may provide a conclusive test of plasma production
at high energies.
However, before the efficacy of the measurement as a test of QGP formation
is proven, the relative importance of other effects must be established.
Although shadowing is important,
the effects should cancel in
ratios of quarkonium states with very similar masses.  
Nuclear absorption should also cancel
if the quarkonium state interacts with nucleons while 
still in a preresonance color octet state \cite{KSoct}.
Although the resonances can interact with comoving secondaries, the 
$p_T$ dependence of these comover interactions is already weak at CERN SPS 
energies \cite{ggj} and expected to be weaker at the LHC \cite{gunv}.  

Thus, if the ratios exhibit a significant $p_T$-dependence 
at large $p_T$ in $AB$ collisions, it will be virtually certain that a quark
gluon plasma was formed.
The precise behavior of the $\psi'/\psi$ and
$\Upsilon'/\Upsilon$ ratios can then be used to strongly constrain
the QGP model parameters.
In particular, the ratios will be very different if only
the $\Upsilon'$ or $\psi'$ is suppressed relative to the case
where all quarkonium states are suppressed.

The $\Upsilon$ rate includes feed down to the $\Upsilon$ 
from $\Upsilon'$, 
$\Upsilon''$ and two sets of $\chi_b$ states and feed down 
to the $\Upsilon'$ from the
$\Upsilon''$ and $\chi_b(2P)$ states.  Thus in the $\Upsilon'/\Upsilon$
ratio, all sources of $\Upsilon'$ and $\Upsilon$, each associated with a 
different suppression factor, must be considered \cite{gunv}:
\begin{equation} \frac{\Upsilon'}{\Upsilon}|_{\rm indirect} \equiv
{\Upsilon'+\chi_b(2P)(\rightarrow \Upsilon')+\Upsilon''(\rightarrow\Upsilon')
\over
\Upsilon+\chi_b(1P,2P)(\rightarrow\Upsilon)+
\Upsilon'(\rightarrow\Upsilon)+\Upsilon''(\rightarrow\Upsilon)}\,.
\label{chain}
\end{equation}
In computing this `indirect' $\Upsilon'/\Upsilon$ ratio it is assumed that
the suppression factor is the same for the $\chi_b(2P)$ and $\chi_b(1P)$ 
states and that identical suppression factors can be used 
for the $\Upsilon'$ and $\Upsilon''$.  The relative 
production and suppression
rates in the color evaporation model, including the $\chi_b$ states,
can be found in Ref.\ \cite{gunv}.
\begin{figure}[htb]
\setlength{\epsfxsize=0.95\textwidth}
\setlength{\epsfysize=0.4\textheight}
\centerline{\epsffile{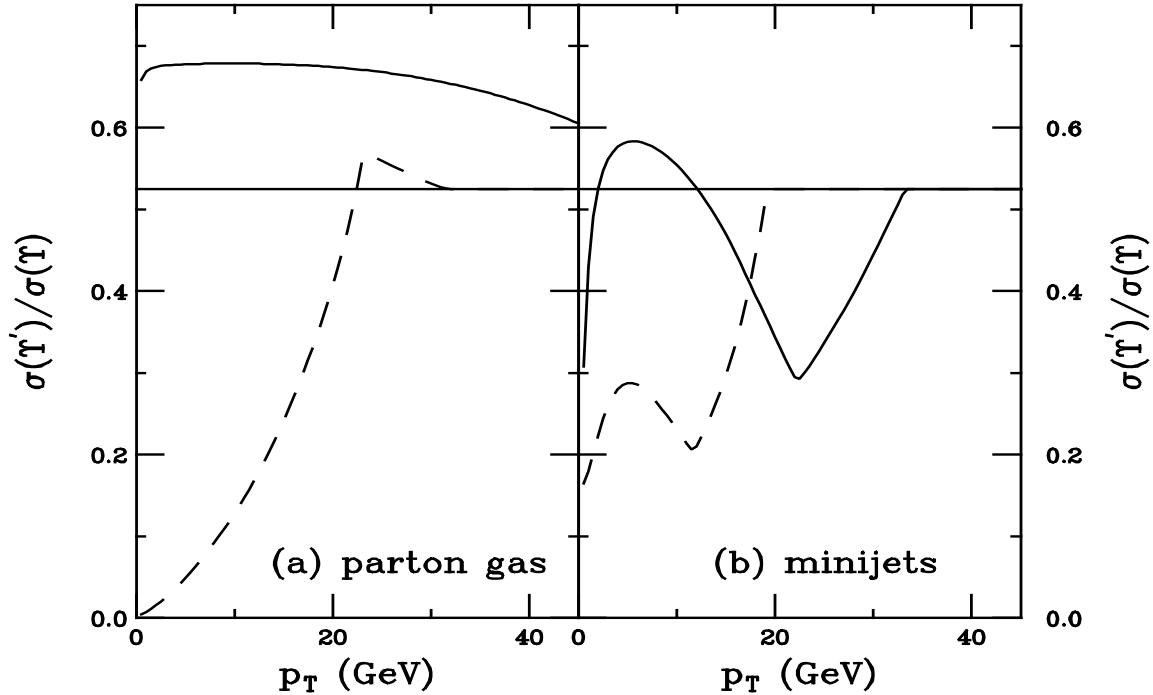}}
\caption[]{The $\Upsilon'/\Upsilon$ ratio computed from 
Eq.~(\protect\ref{chain}) 
is shown for the initial conditions in Table~\protect\ref{tdpttablelhc}
with $R=R_{\rm Pb}$. In (a), parton gas
results are shown for $\mu \propto gT$ (dashed) and $\mu = 4T$ (solid).
In (b) minijet results are given for $\mu = 4T$ without shadowing (solid) and 
with HPC shadowing (dashed).  The horizontal curve represents the $pp$ ratio.
From \protect\cite{cmsNote}.}
\label{figureindirect}
\end{figure}

Figure~\ref{figureindirect} gives the indirect ratios.
In a parton gas assuming $\mu = 4T$, all
the $\Upsilon$ states can be suppressed for $p_T>50$ GeV, producing the rather
flat ratio in the solid curve.
A measurement at the 20\% level is thus needed to distinguish between the $pp$
ratio and the QGP prediction. Substantial systematic errors
in the ratio could make the detection of a deviation quite difficult due
to the slow variation with $p_T$. With the slowly growing
screening mass, $\mu \propto gT$, the direct $\Upsilon$ rate is not suppressed
while the $\Upsilon'$ and $\chi_b$ states
are suppressed.  Under these conditions, the indirect
ratio is less than the $pp$ value until the $\Upsilon'$ is no longer suppressed
and then is slightly enhanced by the $\chi_b$ decays until they also no longer
suffer from plasma effects.  Thus although the indirect ratio is less sensitive
to the plasma, the $\Upsilon'/\Upsilon$ ratios
can significantly constrain plasma models, especially if the quarkonium states
can be measured with sufficient accuracy up to high $p_T$.  

\section{SUMMARY}

In this talk, we have presented only a few of the exciting new physics 
opportunities at the LHC.  For more information on the dedicated heavy ion
experiment ALICE, see the talk of C.~Fabjan \cite{fabjan} and for a taste
of the emerging ultra-peripheral heavy ion program at the LHC, see the talk of
J.~Nystrand \cite{joakim}.  The LHC will certainly turn lead-lead collisions 
into golden data.

\end{document}